\def\rhoav{\overline\rho_{\rm out}}
\def\rhoout{\rho_{\rm out}}
\def\dtwo{d^{\mkern2mu 2}\mkern-1.5mu}
\def\c{{\tt c}}
\def\s{{\tt s}}

\documentclass[prl,superscriptaddress,twocolumn]{revtex4}

\usepackage{bm}

\begin{document}

\title{Classical phase-space descriptions of continuous-variable teleportation}

\author{Carlton M.~Caves}
\email{caves@info.phys.unm.edu} \affiliation{Department of Physics and
Astronomy, University of New Mexico, Albuquerque, NM~87131-1156, USA}
\author{Krzysztof W\'odkiewicz}
\email{wodkiew@fuw.edu.pl} \affiliation{Department of Physics and
Astronomy, University of New Mexico, Albuquerque, NM~87131-1156, USA}
\affiliation{Instytut Fizyki Teoretycznej, Uniwersytet Warszawski,
Warszawa 00--681, Poland}

\date{\today}

\begin{abstract}
The nonnegative Wigner function of all quantum states involved in
teleportation of Gaussian states using the standard continuous-variable
teleportation protocol means that there is a local realistic
phase-space description of the process.  This includes the coherent
states teleported up to now in experiments.  We extend the phase-space
description to teleportation of non-Gaussian states using the standard
protocol and conclude that teleportation of non-Gaussian pure states
with fidelity of $2/3$ is a ``gold standard'' for this kind of
teleportation.
\end{abstract}


\maketitle

Quantum teleportation is a process that can transfer an arbitrary
quantum state from a system held by one party, usually called Alice, to
a system held by a second party, usually called Bob.  The process
requires a pair of systems, shared by Alice and Bob, in an entangled
state---the entangled resource---and a ``small'' amount of classical
information transmitted from Alice to Bob.  Originally proposed for
qubit states \cite{bennett}, teleportation protocols were later
extended to states of a system described by continuous phase-space
variables, such as a massive particle or a mode of an optical field
\cite{vaidman,braunsteinkimble}.  This continuous-variable
teleportation protocol was implemented in an experiment that teleported
a coherent state of an optical-frequency electromagnetic mode with
fidelity $0.58\pm0.02$ \cite{furasawa}.  Two recent experiments have
improved the experimental fidelity of the teleported coherent state to
values of $0.64\pm0.02$ \cite{bowen} and $0.61\pm0.02$ \cite{zhang}.

In the {\it standard continuous-variable teleportation protocol\/}
\cite{braunsteinkimble}, Alice and Bob share an entangled Gaussian
state of two modes, $A$ and $B$, which have annihilation operators $a$
and $b$; this entangled resource is ideally a two-mode squeezed state
\cite{schumaker}.  The state to be teleported is the pure
\cite{purestate} state $\rho=|\psi\rangle\langle\psi|$ of a mode $V$ in
Alice's possession, which has annihilation operator $v$.  The protocol
consists of (i)~Alice's measuring two (commuting) joint quadrature
components of modes $V$ and $A$, specifically the Hermitian real and
imaginary parts of the operator $v+a^\dagger$, (ii)~Alice's
communicating the (complex) result $\xi$ to Bob, and (iii)~Bob's
displacing mode $B$ by $\xi$. The efficacy of the protocol is
quantified by the fidelity between the output state of mode $B$ and the
input state $|\psi\rangle$, averaged over the possible measurement
results.

Experiments to date have teleported only coherent states.  It is
generally believed, though not proved, that teleporting coherent states
with average fidelity ${\cal F}>1/2$ requires an entangled resource
\cite{braunstein00}.  It has thus been argued that teleportation of
coherent states with fidelities above $1/2$ constitutes truly quantum
teleportation \cite{braunstein00,braunstein01}.  Using a variety of
arguments, other workers have contended that ${\cal F}=2/3$ is the
appropriate boundary between classical and quantum teleportation
\cite{ralph,grangier}.  While acknowledging the need for an entangled
resource for teleporting coherent states with fidelity ${\cal F}>1/2$
(indeed, we provide additional evidence), we add a fresh perspective by
investigating whether the entangled resource is used in a way that can
be accounted for by a classical phase-space description. When such a
description exists, it provides a local realistic hidden-variable model
for the teleportation protocol.

Our investigation is motivated by the fact that the coherent states and
the Gaussian entangled resource used in the experiments have {\it
nonnegative\/} Wigner functions \cite{wigner}, which are phase-space
probability distributions that provide a classical description of
measurements of the quadrature components.  A nonnegative Wigner
function does not give a classical description of measurements other
than those of quadrature components; specific such measurements on an
entangled Gaussian state cannot be given a local realistic description
and thus violate Bell inequalities \cite{banaszek}.  Since the standard
protocol uses only quadrature measurements, however, we conclude that
{\it for teleporting coherent states---or any Gaussian state---using
the standard teleportation protocol, the nonnegative Wigner function of
the three modes gives a classical, local realistic description for all
fidelities.}  This means that {\it all the experiments to date---and
any such experiment that teleports coherent states, no matter what
fidelity is achieved---can be accounted for in terms of purely
classical correlations, with no need for a quantum-mechanical
explanation}.

To find situations where the Wigner function does not provide a
classical phase-space description of the standard protocol, we must
look to teleportation of non-Gaussian states, which (for pure states)
have Wigner functions that take on negative values \cite{HP}. To
accommodate non-Gaussian states, we extend our hidden-variable model by
allowing (i)~Alice to substitute a randomly displaced state with
nonnegative Wigner function in place of the non-Gaussian state and
(ii)~Alice and Bob to cheat by teleporting this new ``smeared-out''
state with perfect fidelity. {\it Teleportation of non-Gaussian pure
states with fidelity ${\cal F}\ge2/3$ cannot be accommodated within
this extended hidden-variable model}, thus making a fidelity of 2/3 a
``gold standard'' for teleportation of non-Gaussian pure~states.

We begin with a brief Wigner-function-based review of the teleportation
protocol. The state $\rho_{AB}$ of modes $A$ and $B$ has Wigner
function $W_{AB}(\alpha,\beta)$, which is a quasidistribution for the
c-number complex amplitudes $\alpha$ and $\beta$ corresponding to the
annihilation operators $a$ and $b$ \cite{garzoll}.  In the standard
protocol, $W_{AB}(\alpha,\beta)$ is a Gaussian, but for the present, we
allow it to be a general Wigner function.  The (pure \cite{purestate})
state $\rho=|\psi\rangle\langle\psi|$ of mode $V$ has Wigner function
$W_\rho(\nu)$, where $\nu$ is the c-number complex amplitude
corresponding to annihilation operator~$v$.  The overall Wigner
function of the three modes is $W_\rho(\nu)W_{AB}(\alpha,\beta)$.  The
state of mode~$B$ after a measurement of $v+a^\dagger$ that yields
result $\xi$ has Wigner function
\begin{equation}
W'(\beta|\,\xi)\!=\! {1\over p(\xi)}\!\int\!\dtwo\nu\,\dtwo\alpha\,
\delta(\nu+\alpha^*-\xi)W_\rho(\nu)W_{AB}(\alpha,\beta)\;,
\end{equation}
where
\begin{equation}
p(\xi)= \int \dtwo\nu\,\dtwo\alpha\,\dtwo\beta\,
\delta(\nu+\alpha^*-\xi)W_\rho(\nu)W_{AB}(\alpha,\beta)
\end{equation}
is the probability to obtain result $\xi$.

Having received result $\xi$ from Alice, Bob displaces the complex
amplitude of mode $B$ by $\xi$, yielding a state $\rhoout(\xi)$ with
Wigner function $W_{\rm out}(\beta|\xi)=W'(\beta-\xi|\xi)$.  The
fidelity of this output state and the input state is
$F(\xi)=\langle\psi|\rhoout(\xi)|\psi\rangle$.  We are interested in
the average of this fidelity over all measurement results, ${\cal
F}=\int\dtwo\xi\,p(\xi)F(\xi)= \langle\psi|\,\rhoav|\psi\rangle$, where
$\rhoav=\int\dtwo\xi\,p(\xi)\rhoout(\xi)$ is the average output state,
having Wigner function
\begin{equation}
\label{Wout}
W_{\rhoav}(\beta)\!=\! \!\int\!\dtwo\xi\,p(\xi)W_{\rm
out}(\beta|\,\xi)\!=\! \!\int\!\dtwo\nu\,G(\nu)W_\rho(\beta-\nu)\;.
\end{equation}
Here
\begin{equation}
\label{defG}
G(\nu)= \int\dtwo\alpha\,\dtwo\beta\,
\delta(\beta+\alpha^*-\nu)W_{AB}(\alpha,\beta)
\end{equation}
is the (nonnegative) probability to obtain result $\nu$ in a
measurement of $b+a^{\dagger}$ on modes $A$ and $B$.
Equation~(\ref{Wout}) shows that the average output state is a mixture
of displaced input states,
\begin{equation}
\rhoav= \int\dtwo\nu\,G(\nu)D(\nu)\rho D^\dagger(\nu)\;,
\label{rhoav}
\end{equation}
where $D(\nu)$ is the displacement operator.

We can now write the average output fidelity in two complementary
forms,
\begin{eqnarray}
{\cal F}&=&
\int\dtwo\nu\,G(\nu)|C_\rho(\nu)|^2\nonumber\\
&=&\pi\int\dtwo\beta\,\dtwo\nu\,G(\beta-\nu)W_\rho(\beta)W_\rho(\nu)\;,
\label{fidelity1}
\end{eqnarray}
where
\begin{equation}
C_\rho(\nu)=\langle\psi|D(\nu)|\psi\rangle
=\int\dtwo\mu\,W_\rho(\mu)e^{\nu\mu^*-\nu^*\mu}\;,
\label{FT}
\end{equation}
the symmetrically ordered (Wigner-Weyl) characteristic function of the
input state \cite{cahillglauber}, is the Fourier transform of the
Wigner function.  The first form in Eq.~(\ref{fidelity1}) comes
directly from Eq.~(\ref{rhoav}), and the second from from writing the
fidelity as an overlap of the Wigner functions for the input and
average output states.  The effect of the initial state of modes $A$
and $B$ on the average fidelity is contained wholly in the marginal
distribution $G(\nu)$.  High-fidelity teleportation occurs when
$G(\nu)$ is very narrow, i.e., when the quadrature components contained
in $b+a^\dagger$ are sharp, expressing a particular kind of correlation
between modes $A$ and $B$.  Using the Fourier transform~(\ref{FT}), we
can derive two other, equivalent forms for the average fidelity,
\begin{eqnarray}
{\cal F}
&=&\pi\int\dtwo\beta\,\dtwo\nu\,\tilde G(\beta-\nu)
W_\rho(\beta)W_\rho(\nu)\nonumber\\
&=&\int\dtwo\nu\,\tilde G(\nu)|C_\rho(\nu)|^2\;,
\label{fidelity2}
\end{eqnarray}
where
\begin{equation}
\tilde G(\nu)=\int\dtwo\mu\,G(\mu)e^{\nu\mu^*-\nu^*\mu}
\end{equation}
is the Fourier transform of $G(\mu)$.

Before proceeding to the standard protocol and our hidden-variable
models, we pause here to demonstrate the one technical result we need.
We wish to find the maximum value of the integral
\begin{equation}
\label{I1}
I=\int\dtwo\alpha\,\dtwo\beta\, e^{-t|\alpha-\beta|^2/2}
W_{AB}(\alpha,\beta)\;, \quad t\ge0,
\end{equation}
over the Wigner function $W_{AB}(\alpha,\beta)$ of a joint state
$\rho_{AB}$ of modes $A$ and $B$.  Introducing annihilation operators
$c=(a+b)/\sqrt2$ and $d=(a-b)/\sqrt2$, with corresponding c-number
variables $\gamma$ and $\delta$, we can rewrite $I$ as
\begin{equation}
\label{I2}
I=\int\dtwo\gamma\,\dtwo\delta\,e^{-t|\delta|^2}W_{CD}(\gamma,\delta)
=\int\dtwo\delta\,e^{-t|\delta|^2}W_D(\delta)\;,
\end{equation}
where $W_{CD}(\gamma,\delta)=W_{AB}(\alpha,\beta)$ is the Wigner
function written in terms of modes $C$ and $D$, and $W_D(\delta)$ is
the Wigner function for mode~$D$ alone.  The integral can now be
thought of as the expectation value, ${\rm tr}(A_t\rho_D)={\rm
tr}(A_t\rho_{AB})$, of the $D$-mode operator $A_t$ whose symmetrically
ordered associated function \cite{cahillglauber} is $e^{-t|\delta|^2}$,
this operator being $A_t=(1+t/2)^{-1}[(1-t/2)/(1+t/2)]^{d^\dagger d}$.

The integral $I$ being the expectation value of $A_t$, $I$ is bounded
above by the largest eigenvalue of $A_t$.  Since $A_t$ is diagonal in
the number-state basis, with eigenvalues that decrease in magnitude
with the number of quanta, we have
\begin{equation}
I_{\rm max}=\hbox{(largest eigenvalue of $A_t$)}={1\over1+t/2}\;,
\label{Imax}
\end{equation}
with the maximum achieved if and only if $\rho_{AB}$ is the vacuum
state for mode~$D$.

For the case that $\rho_{AB}$ is a pure product state,
$|\Psi\rangle=|\psi_A\rangle\otimes|\psi_B\rangle$, which turns out to
be the case of interest here, the condition for achieving the maximum
becomes $d|\Psi\rangle=0$ or, equivalently,
$a|\psi_A\rangle\otimes|\psi_B\rangle=|\psi_A\rangle\otimes\,b|\psi_B\rangle$,
from which it follows that $a|\psi_A\rangle=|\psi_A
\rangle\langle\psi_B|b|\psi_B\rangle$ and $b|\psi_B\rangle=|\psi_B
\rangle\langle\psi_A|a|\psi_A\rangle$, implying that $|\psi_A\rangle$
and $|\psi_B\rangle$ are identical coherent states.  Thus the only pure
product states that achieve the maximum in Eq.~(\ref{Imax}) are
products of identical coherent states.

We can use Eq.~(\ref{Imax}) to get one interesting result immediately:
{\it The maximum average fidelity for teleporting a coherent state
using the standard protocol, but with a separable state for modes $A$
and $B$, is $1/2$}.  To show this, suppose first that modes $A$ and $B$
are in a pure product state, with factorizable Wigner function
$W_A(\alpha)W_B(\beta)$.  The characteristic function for any coherent
state satisfies $|C_{\rm coh}(\nu)|^2=e^{-|\nu|^2}$, so we can use
Eq.~(\ref{defG}) and the first form in Eq.~(\ref{fidelity1}) to write
the average fidelity as
\begin{equation}
{\cal F}= \int\dtwo\alpha\,\dtwo\beta\,
e^{-|\alpha-\beta|^2}W_A(-\alpha^*)W_B(\beta)\;.
\end{equation}
Here $W_A(-\alpha^*)$ is the Wigner function for the time-reversed,
parity-inverted state of mode~$A$.  The $t=2$ general
bound~(\ref{Imax}) implies ${\cal F}\le1/2$, with equality if and only
if mode $A$ is in a coherent state $|\alpha\rangle$ and mode $B$ is in
the time-reversed, parity-inverted coherent state
$|\mathord{-}\alpha^*\rangle$.

Now suppose modes $A$ and $B$ are initially in a separable state, thus
having a pure product-state ensemble decomposition.  The fidelity is
the average over the pure product-state ensemble, which shows that the
fidelity is still bounded above by $1/2$, with equality if and only if
the separable state is a mixture of product states of the form
$|\alpha\rangle\otimes|\mathord{-}\alpha^*\rangle$.  This does {\it
not\/} show that $1/2$ is the maximum fidelity for coherent-state
teleportation in the absence of entanglement, since the result applies
only to the standard protocol, but it is an additional piece of
evidence, distinct from the results reported in
Ref.~\cite{braunstein01}.

We now take up again our analysis of the standard teleportation
protocol, assuming that modes $A$ and $B$ are in a Gaussian state with
Wigner function
\begin{equation}
\label{Wepr}
W_{AB}(\alpha,\beta)= {4(\c^2-\s^2)\over\pi^2}\,
e^{-2\c(|\alpha|^2+|\beta|^2) -2\s(\alpha\beta+\alpha^*\beta^*)}\;,
\end{equation}
where $\c$ and $\s$ satisfy $|\s|<\c\le\sqrt{1+\s^2}$.  This state
is pure if and only if $\c=\sqrt{1+\s^2}$, in which case the state
becomes a two-mode squeezed state with $\c=\cosh2r$ and
$\s=\sinh2r$, where $r$ is the squeeze parameter \cite{schumaker}.
The state~(\ref{Wepr}) is separable if and only if $\c\le1-|\s|$
\cite{tutorial}.

For the Wigner function~(\ref{Wepr}), the distribution~(\ref{defG}) is
a Gaussian,
\begin{equation}
G(\nu)={\c+\s\over\pi}\,e^{-(\c+\s)|\nu|^2}= {2\over\pi
t}\,e^{-2|\nu|^2/t}\;,
\end{equation}
where $t\equiv2/(\c+\s)$ is the single parameter needed to characterize
the fidelity that can be achieved with this entanglement resource.  The
Wigner function~(\ref{Wout}) of the average output state is the
$(s=-t)$-ordered quasidistribution, $W_\rho^{(s)}(\nu)$
\cite{cahillglauber}, of the input state:
\begin{equation}
W_{\rhoav}(\beta)=
{2\over\pi t}\int\dtwo\nu\,e^{-2|\beta-\nu|^2/t}W_\rho(\nu)
=W_\rho^{(s)}(\nu) \;.
\end{equation}
For $t=0$, $G(\nu)$ is a $\delta$-function, and the output state is
identical to the input state (perfect teleportation).  For $t=1$, the
Wigner function of the average output state is the Husimi $Q$
distribution of the input state, i.e.,
$W_{\rhoav}(\beta)=W_\rho^{(-1)}(\beta)=Q_\rho(\beta)
=\langle\beta|\rho|\beta\rangle/\pi$.

For $0\le t<2$ ($\c>1-\s$), the state~(\ref{Wepr}) is entangled, with
the right sort of correlations for this protocol, these correlations
decreasing as $t$ increases.  At $t=2$, the state passes through the
separability boundary $\c+\s=1$, and for $t\ge2$, the state either is
separable ($\c\le1-|\s|$) or, though entangled ($\c>1+\s$), has the
wrong sort of correlations for this protocol.

The average fidelity of Eqs.~(\ref{fidelity1}) and (\ref{fidelity2})
now becomes
\begin{eqnarray}
{\cal F}_\rho(t)
&=&{2\over\pi t}\int\dtwo\nu\,e^{-2|\nu|^2/t}|C_\rho(\nu)|^2\nonumber\\
&=&{2\over t}\int\dtwo\beta\,\dtwo\nu\,e^{-2|\beta-\nu|^2/t}
W_\rho(\beta)W_\rho(\nu)\nonumber\\
&=&\int\dtwo\beta\,\dtwo\nu\,e^{-t|\beta-\nu|^2/2}
W_\rho(\beta)W_\rho(\nu)\nonumber\\
&=&{1\over\pi}\int\dtwo\nu\,e^{-t|\nu|^2/2}|C_\rho(\nu)|^2
\;.
\label{fidelity3}
\end{eqnarray}
The second form is the overlap of the Wigner function and the
$s$-ordered quasidistribution for the input state.  The first two
derivatives of the last form show that ${\cal F}_\rho(t)$ is a strictly
decreasing, strictly concave function of $t$.  These forms also show
that the average fidelity obeys the scaling relation ${\cal F}_\rho(t)
=2{\cal F}_\rho(4/t)/t$, which again draws attention to the
separability boundary at $t=2$.

Given that $|C_{\rm coh}(\nu)|^2=e^{-|\nu|^2}$, the average fidelity
for teleporting a coherent state is ${\cal F}_{\rm
coh}(t)=(1+t/2)^{-1}$.  For number states $|n\rangle$, whose Wigner
functions take on negative values (except for $n=0$), we have obtained
an analytic formula for a generating function
\begin{eqnarray}
\mathcal{F}(\lambda,t)\!&=&\!
\sum_{n=0}^{\infty}\lambda^n\mathcal{F}_{|n\rangle\langle n|}(t)\\
\!&=&\!\frac{1}{\sqrt{(1+t/2)^2-2\lambda(1+t^2/4)
+\lambda^2(1-t/2)^2}}\;.\nonumber
\end{eqnarray}
The resulting fidelity for teleporting a number state is
\begin{equation}
\mathcal{F}_{|n\rangle\langle n|}(t)= \frac{(1-t/2)^n}{(1+t/2)^{n+1}}\,
P_n\!\left(\frac{1+t^2/4}{1-t^2/4}\right)\;,
\end{equation}
where $P_n(x)$ is a Legendre polynomial.  This gives
$\mathcal{F}_{|n\rangle\langle n|}(1)=2P_n(5/3)/3^{n+1}$ and
$\mathcal{F}_{|n\rangle\langle n|}(2)=(2n)!/2^{2n+1}(n!)^2$.  Another
example of a state with a Wigner function that takes on negative values
is the superposition $|\psi \rangle=(|0\rangle +|1\rangle)/\sqrt2$,
which has teleportation fidelity
$\mathcal{F}_{|\psi\rangle\langle\psi|}(t)= (1+3t/4+t^2/4)/(1+t/2)^3$.

We now return to our Wigner-function-based discussion of
hidden-variable models for teleportation.  For Gaussian input states
and for the two-mode entangled resource~(\ref{Wepr}), all the Wigner
functions are nonnegative \cite{wigner}, so they provide a classical
phase-space description---and hence a local hidden-variable
description---for this kind of teleportation, no matter what fidelity
is achieved.  The hidden variables are the quadrature components of all
the modes, and the overall Wigner function is a probability
distribution for these hidden variables.

All non-Gaussian input pure states have Wigner functions that take on
negative values (Hudson-Piquet theorem \cite{HP}) and thus cannot be
incorporated in the simple hidden-variable model.  To see what can be
achieved within a classical phase-space description, suppose that
before performing the teleportation protocol, Alice ``kicks'' the input
state $\rho$ randomly in phase space.  The random kick is described by
a Gaussian so that the average state after the kick is
\begin{equation}
\rho'= {2\over\pi t}\int\dtwo\nu\,e^{-2|\nu|^2/t}
D(\nu)\rho D^\dagger(\nu)\;.
\end{equation}
We choose the kicking strength $t$ to be the {\it minimum\/} value
necessary to make $\rho'$ have a nonnegative Wigner function, thus
giving $\rho'$ a classical phase-space description and allowing it to
be incorporated it within our hidden-variable model.  For all
non-Gaussian states, this minimum kicking strength is one vacuum unit,
i.e., $t=1$ \cite{lutkenhaus}, implying that $\rho'$ is the state whose
Wigner function, $W_{\rho'}(\nu)=W_\rho^{(-1)}(\nu)=Q_\rho(\nu)$, is
the $Q$ function of the original state $\rho$.  Further suppose that
Alice and Bob cheat by teleporting $\rho'$ with perfect fidelity.  Then
the fidelity of the overall process is the overlap of the Wigner and
$Q$ functions of $\rho$, i.e., the $t=1$ fidelity~(\ref{fidelity3}) of
the standard protocol.  Notice that bigger kicks ($t>1$) would give
smaller fidelity, making clear why we choose the smallest kicking
strength consistent with giving $\rho'$ a nonnegative Wigner function.

These considerations, coupled with wanting to know the maximum
teleportation fidelity for a given entangled resource $t$, motivate us
to find the maximum value of the average fidelity $\mathcal{F}_\rho(t)$
over all input pure states $\rho$, $t=1$ being the value relevant for
our hidden-variable model.  The task can be restated as finding the
pure state $\rho$ that maximizes the overlap of the Wigner function and
the $s$-ordered quasidistribution $W_\rho^{(s)}(\nu)$.  We apply the
bound~(\ref{Imax}) to the third form of the average
fidelity~(\ref{fidelity3}), in this case maximizing over pure product
states $\rho\otimes\rho$.  The resulting maximum is ${\cal F}_{\rm
max}(t)=(1+t/2)^{-1}$, achieved if and only if $\rho$ is a coherent
state.

Each non-Gaussian input state $\rho$ has its own threshhold fidelity,
${\cal F}_\rho(1)<{\cal F}_{\rm max}(1)=2/3$, below which its
teleportation can be accommodated within our extended phase-space
hidden-variable model and above which it cannot.  Thus teleportation of
a non-Gaussian state with fidelity exceeding ${\cal F}_\rho(1)$ is
required to rule out an explanation in terms of classical phase-space
correlations.  A fidelity of $2/3$ emerges as a ``gold standard'' for
continuous-variable teleportation in the sense that teleportation of
{\it any\/} non-Gaussian pure state with ${\cal F}\ge2/3$ cannot be
fitted within our extended hidden-variable model. This conclusion
applies only to our phase-space-based hidden-variable model; we have
not shown that there is no local hidden-variable model that can
accommodate teleportation fidelities of 2/3 or above.

This work was supported in part by US Office of Naval Research Grant
No.~N00014-03-1-0426 (CMC) and by KBN Grant No.~2PO3B\thinspace02123
(KW).


\begin{thebibliography}{99}
\bibitem{purestate}
We restrict attention to teleporting pure states.
\end{thebibliography}
\end{document}